\documentclass[a4paper,10pt]{article}
\usepackage{amssymb}
\usepackage{amsmath}
\usepackage{graphicx}

\title{The Hubbard model in the strong coupling theory at arbitrary filling}

\author{A. Sherman\\[1ex]
{\it Institute of Physics, University of Tartu, Ravila 14c, 50411 Tartu, Estonia}}

\begin{document}

\maketitle

\begin{abstract}
Equations for the electron Green's function of the two-dimensional Hubbard model, derived using the strong coupling diagram technique, are self-consistently solved for different electron concentrations $n$ and tight-binding dispersions. Comparison of spectral functions calculated for the ratio of Hubbard repulsion to the nearest neighbor hopping $U/t=8$ with Monte Carlo data shows not only qualitative, but in some cases quantitative agreement in position of maxima. General spectral shapes, their evolution with momentum and filling in the wide range $0.7\lesssim n\leq 1$ are also similar. At half-filling and for the next nearest neighbor hopping constant $t'=-0.3t$ the Mott transition occurs at $U_c\approx 7\Delta/8$, where $\Delta$ is the initial bandwidth. This value is close to those obtained in the cases of the semi-elliptical density of states and for $t'=0$. In the case $U=8t$ and $t'=-0.3t$ the Mott gap reaches maximum width at $n=1.04$, and it is larger than that at $t'=0$ for half-filling. In all considered cases positions of spectral maxima are close to those in the Hubbard-I approximation.
\end{abstract}

\section{Introduction}
The fermionic Hubbard model is one of the main models used for the description of strong electron correlations in crystals such as cuprate perovskites and heavy fermion compounds. The strong coupling diagram technique \cite{Vladimir,Metzner,Craco,Pairault,Sherman06} is one of the approximate approaches used for investigating the model. The method is based on the serial expansion in powers of the electron kinetic energy. The elements of the arising diagram technique are on-site cumulants of electron creation and annihilation operators and hopping lines connecting cumulants on different sites. As in the diagram technique with the expansion in powers of an interaction, in the considered approach the linked-cluster theorem allows one to discard disconnected diagrams and to carry out partial summations in remaining connected diagrams. As a consequence the electron Green's function is expressed in the form of the Larkin equation containing the initial electron dispersion and the irreducible part -- the sum of all irreducible diagrams without external ends. In spite of the fact that by the construction the approach is intended for the case of strong coupling, when the Hubbard repulsion $U$ is approximately equal to or larger than the initial bandwidth $\Delta$, it gives the correct result in the limit $U\rightarrow 0$. Hence the approach provides an interpolation between the limits of weak and strong correlations.

Using this method, in recent work \cite{Sherman15} equations for the electron Green's function were obtained by keeping terms of the lowest two orders in the irreducible part of the Larkin equation. Self-consistent calculations performed for the semi-elliptical density of states (DOS) showed that at half-filling the approximation describes the Mott transition, which occurs at $U_c=\sqrt{3}\Delta/2$. This value coincides with the critical repulsion obtained for the same DOS in the Hubbard-III approximation \cite{Hubbard64}. In \cite{Sherman14} the same method was applied to the half-filled two-dimensional (2D) Hubbard model with nearest neighbor form of the kinetic energy (the $t$-$U$ model). In this case the Mott transition is observed at $U_c\approx 7\Delta/8$ -- the value, which is very close to that for the semi-elliptical DOS. The used approach allows one to describe not only the Mott transition, but also another important feature of the model -- spectral continua, which lead to finite widths of spectral maxima. Spectral functions, DOS and momentum distributions calculated in the $t$-$U$ model were in qualitative and in some cases quantitative agreement with Monte Carlo data.

In the present work the equations derived in \cite{Sherman15} are self-consistently solved for different electron concentration $n$ and for the kinetic energy containing nearest and next nearest neighbor hopping terms (the $t$-$t'$-$U$ model). Similar dispersions are frequently used for the description of hole-doped cuprates with the ratio of the next nearest neighbor to nearest neighbor hopping constants close to $t'/t=-0.3$ (see, e.g., \cite{Macridin}). At half-filling and for this ratio of $t'/t$ the Mott transition is observed again at $U_c\approx 7\Delta/8$, close to the values mentioned above. Comparison of calculated spectral functions with available Monte Carlo results for the case $U=8t$ and $t'=0$ shows not only qualitative, but in some cases quantitative agreement in positions of maxima. General spectral shapes, their evolution with wave vector and concentration in the wide range $0.7\lesssim n\leq 1$ are also close. These facts and relative simplicity of calculations give promise that the considered approach can be applied for investigating multi-band generalizations of the Hubbard model, for example, such as used for the description of iron pnictides or transition-metal oxides. It was found that the behaviour of the Mott gap with changing $n$ is different in the $t$-$U$ and $t$-$t'$-$U$ models. In all considered cases positions of maxima in spectral functions are close to the frequencies of $\delta$-function peaks in the Hubbard-I approximation \cite{Hubbard63}.

\section{Main formulas}
In this section some equations derived in \cite{Sherman15} are reproduced and converted to the form, which is convenient for calculations. The electron Green's function
\begin{equation}\label{Green}
G({\bf n'\tau',l\tau})=\langle{\cal T}\bar{a}_{\bf n'\sigma}(\tau')
a_{\bf n\sigma}(\tau)\rangle
\end{equation}
is considered, where the angular brackets denote the statistical averaging with the Hubbard Hamiltonian
\begin{equation}\label{Hamiltonian}
H=\sum_{\bf nn'\sigma}t_{\bf nn'}a^\dagger_{\bf n\sigma}a_{\bf n'\sigma} +\frac{U}{2}\sum_{\bf n\sigma}n_{\bf n\sigma}n_{\bf n,-\sigma}
-\mu\sum_{\bf n\sigma}n_{\bf n\sigma},
\end{equation}
$t_{\bf nn'}$ is the hopping constants, the operator $a^\dagger_{\bf n\sigma}$ crea\-tes an electron on the site {\bf n} of the 2D square lattice with the spin projection $\sigma=\pm 1$, the electron number operator $n_{\bf n\sigma}=a^\dagger_{\bf n\sigma}a_{\bf n\sigma}$, $\mu$ is the chemical potential, ${\cal T}$ is the time-ordering operator which
arranges operators from right to left in ascending order of times $\tau$,
$a_{\bf n\sigma}(\tau)=\exp(H\tau)a_{\bf n\sigma}\exp(-H\tau)$ and $\bar{a}_{\bf n\sigma}(\tau)=\exp(H\tau)a^\dagger_{\bf n\sigma}\exp(-H\tau)$. Green's function (\ref{Green}) does not depend on the spin projection, and it was omitted in the function notation.

In the strong coupling diagram technique, after the summation of all diagrams the Fourier transform of Green's function (\ref{Green}) acquires the form of the Larkin equation
\begin{equation}\label{Larkin}
G({\bf k},i\omega_l)=\frac{K({\bf k},i\omega_l)}{1-t_{\bf k}K({\bf k},i\omega_l)},
\end{equation}
where ${\bf k}$ is the 2D wave vector, $\omega_l=(2l+1)\pi T$ is the Matsubara frequency with the temperature $T$, $t_{\bf k}=\sum_{\bf n}\exp[i{\bf k(n-n')}]t_{\bf nn'}$ and $K({\bf k},i\omega_l)$ is the irreducible part -- the sum of all irreducible diagrams without external ends. A diagram is said to be an irreducible one if it cannot be divided into two disconnected parts by cutting some hopping line $t_{\bf nn'}$.

In this work $K({\bf k},i\omega_l)$ is approximated by the sum of two terms of the lowest orders in powers of $t_{\bf k}$. These terms contain cumulants of the first and second orders. The term with the second-order cumulant includes also a hopping-line loop. Using the possibility of the partial summation of diagrams we transform this bare hopping line into the dressed one,
\begin{eqnarray}\label{hopping}
&&\theta({\bf n}\tau,{\bf n'}\tau')=t_{\bf nn'}\delta(\tau-\tau')+\sum_{\bf mm'}t_{\bf nm}\nonumber\\
&&\quad\quad\quad\times\int_0^\beta d\upsilon\, K({\bf m}\tau,{\bf m'}\upsilon)\, \theta({\bf m'}\upsilon,{\bf n'}\tau'),
\end{eqnarray}
where $\beta=1/T$. With this substitution, for a chemical potential in the range
\begin{equation}\label{mu_range}
\mu\gg T,\quad U-\mu\gg T
\end{equation}
the expression for the $K({\bf k},i\omega_l)$ reads
\begin{eqnarray}\label{K_gen}
&&K(i\omega_l)=\frac{1}{2}\left[g_{01}(i\omega_l)+g_{12}(i\omega_l)\right]\nonumber\\
&&\quad+\frac{3}{4}F^2(i\omega_l)\phi(i\omega_l) -\frac{s_1}{2}F(i\omega_l)-\frac{s_2}{2}J(i\omega_l),
\end{eqnarray}
where
\begin{eqnarray}\label{variables}
&&g_{01}(i\omega_l)=\left(i\omega_l+\mu\right)^{-1},\, g_{12}(i\omega_l)=\left(i\omega_l+\mu-U\right)^{-1},\nonumber\\ &&F(i\omega_l)=g_{01}(i\omega_l)-g_{12}(i\omega_l),\nonumber\\
&&J(i\omega_l)=g^2_{01}(i\omega_l)-F(i\omega_l)g_{12}(i\omega_l),\\
&&s_1=T\sum_l J(i\omega_l)\phi(i\omega_l),\quad
s_2=T\sum_l F(i\omega_l)\phi(i\omega_l),\nonumber\\
&&\phi(i\omega_l)=N^{-1}\sum_{\bf k}t^2_{\bf k}G({\bf k},i\omega_l)\nonumber
\end{eqnarray}
and $N$ is the number of sites. We chose $t_{\bf nn}=0$ and used this relation in deriving (\ref{K_gen}). Equations (\ref{Larkin}), (\ref{K_gen}) and (\ref{variables}) form the closed set of equation for calculating Green's function (\ref{Green}). Notice that irreducible part (\ref{K_gen}) does not depend on momentum as the approximation does not take into account the interactions of electrons with the magnetic ordering, spin and charge fluctuations. In the used approach these interactions are described by the sum of diagrams containing ladders, which correspond to the dynamic spin and charge susceptibilities \cite{Sherman07}.

For further consideration we perform the analytic continuation to real frequencies $\omega$ and transform the sums $s_1$ and $s_2$ to the form more convenient for calculations using the spectral representations, Poisson summation formulas and inequalities (\ref{mu_range}),
\begin{eqnarray}\label{s1_2}
s_1&=&-\int_{-\infty}^0\frac{2U-\omega-\mu}{U(\omega+\mu-U)^2}\,{\rm Im}\phi(\omega)\frac{d\omega}{\pi}\nonumber\\
&&+\int_0^\infty\frac{U+\omega+\mu}{U(\omega+\mu)^2}\,{\rm Im}\phi(\omega)\frac{d\omega}{\pi},\\
s_2&=&\int_{-\infty}^0\frac{\rm Im\phi(\omega)}{\omega+\mu-U}\frac{d\omega}{\pi}
+\int_0^\infty\frac{\rm Im\phi(\omega)}{\omega+\mu}\frac{d\omega}{\pi}.\nonumber
\end{eqnarray}
From (\ref{Larkin}), (\ref{K_gen}) and (\ref{variables}) we get
\begin{eqnarray}\label{phi}
\phi(\omega)&=&(\omega+\mu)^2(\omega+\mu-U)^2N^{-1}\sum_{\bf k}t_{\bf k}\nonumber\\
&&\times\bigg\{(\omega+\mu)^2(\omega+\mu-U)^2\nonumber\\
&&-t_{\bf k}\bigg[\bigg(\omega+\mu +\frac{U}{2}(s_1-1)\bigg)(\omega+\mu)\nonumber\\
&&\times(\omega+\mu-U)-\frac{s_2}{2}\bigg((\omega+\mu-U)^2\nonumber\\
&&+U(\omega+\mu)\bigg)+\frac{3}{4}U^2\phi(\omega)\bigg]\bigg\}^{-1}.
\end{eqnarray}
A self-consistent solution $\phi(\omega)$ of (\ref{s1_2}) and (\ref{phi}) is used in (\ref{K_gen}) and (\ref{Larkin}) for calculating Green's function.
\begin{figure}[t]
\centerline{\resizebox{0.5\columnwidth}{!}{\includegraphics{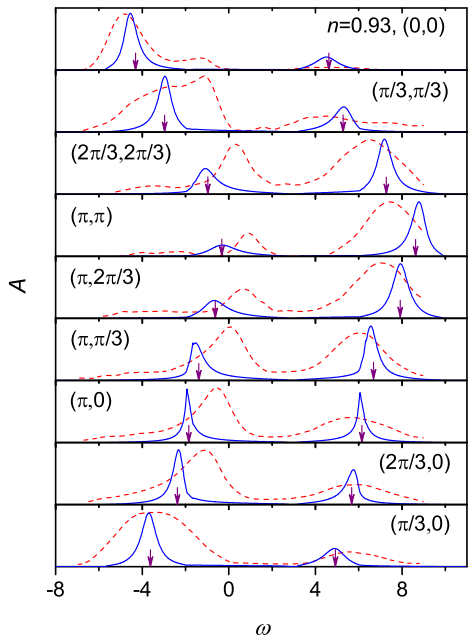}}}
\caption{Spectral functions in the $t$-$U$ model for $U=8t$, $n=0.93$, a 6$\times$6 lattice and momenta shown in the panels. Blue solid lines are our results, red dashed lines are Monte Carlo simulations for $T=0.5t$ \protect\cite{Groeber}, arrows indicate locations of $\delta$-function peaks of the Hubbard-I approximation.} \label{Fig1}
\end{figure}
\begin{figure}[t]
\centerline{\resizebox{0.5\columnwidth}{!}{\includegraphics{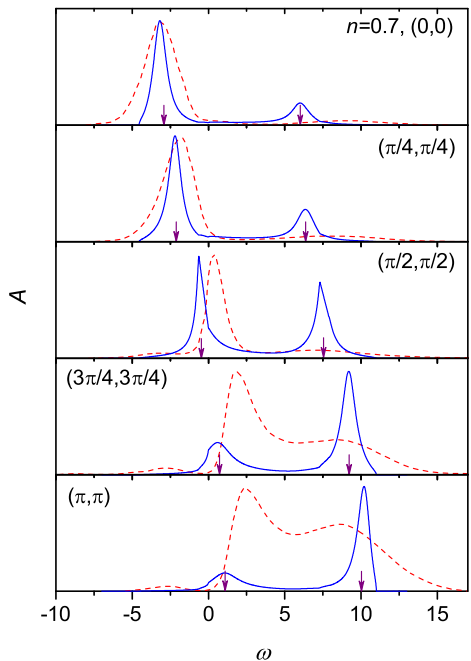}}}
\caption{Same as in Fig.~\protect\ref{Fig1}, but for $n=0.7$ and a 8$\times$8 lattice. The Monte Carlo data for $T=0.5t$ are taken from \protect\cite{Haas}.} \label{Fig2}
\end{figure}

As follows from (\ref{K_gen})--(\ref{phi}), frequencies $\omega=-\mu$ and $U-\mu$ are somehow singled out. At half-filling and in the case of a DOS, which is symmetric with respect to $\mu=U/2$, the sums $s_1$ and $s_2$ vanish. As follows from (\ref{phi}), in this case $\phi(\omega)$ behaves near $\omega=\pm U/2$ as (see also \cite{Sherman14})
$$\phi\bigg(\omega\approx\pm\frac{U}{2}\bigg)\approx -\frac{1}{3}\bigg(\omega\pm\frac{U}{2}\bigg)-i\frac{\sqrt{11}}{3}\bigg|\omega\pm\frac{U}{2}\bigg|.$$
Substituting this result into (\ref{K_gen}) we see that ${\rm Im}\,K(\omega)$ diverges as $(\omega\pm U/2)^{-1}$. This fact presents a problem because up to a multiplier ${\rm Im}\,K(\omega)$ is a spectral function for momenta meeting the condition $t_{\bf k}=0$ [see (\ref{Larkin})], which due to the divergence does not satisfy neither the normalization condition
\begin{equation}\label{norm}
\int_{-\infty}^\infty{\rm Im}\,K(\omega)\,d\omega=-\pi
\end{equation}
nor the Kramers-Kronig relation
\begin{equation}\label{kkr}
{\rm Re}\,K(\omega)=\int_{-\infty}^\infty\frac{{\rm Im}\,K(\omega')}{\omega'-\omega}\frac{d\omega}{\pi}.
\end{equation}
To overcome this difficulty in \cite{Sherman14} we introduced an artificial broadening. However, if an initial DOS is not symmetric (as for the case $t'\neq 0$) and/or $n\neq 1$ the divergence disappears. Indeed, in this case $s_1$ and $s_2$ in (\ref{s1_2}) are nonzero, and from (\ref{phi}) we get near $\omega=-\mu$ and $U-\mu$
\begin{eqnarray*}
&&\phi(\omega\approx-\mu)\approx\frac{2}{s_2}(\omega+\mu)^2, \\
&&\phi(\omega\approx U-\mu)\approx\frac{2}{s_2}(\omega+\mu-U)^2.
\end{eqnarray*}
Here $\phi(\omega)$ is purely real and, therefore, ${\rm Im}\,K=0$. This poses another difficulty: as follows from calculations, for $s_2\neq 0$ narrow gaps with nearly vertical walls appear in ${\rm Im}\,K(\omega)$  around $\omega=-\mu$ and $U-\mu$. Thanks to their narrowness and shape, these unphysical gaps can be easily eliminated by a linear interpolation between tops of the walls.

\section{Results and discussion}
Results of this section were obtained by the self-con\-si\-s\-tent solution of (\ref{s1_2}) and (\ref{phi}) for a given $\mu$ in the $t$-$U$ or $t$-$t'$-$U$ models. The solution was found by iteration. Such obtained $\phi(\omega)$, $s_1$ and $s_2$ were used for calculating ${\rm Im}\,K(\omega)$ from (\ref{K_gen}). After removing the gaps near $\omega=-\mu$ and $U-\mu$ with the interpolation and normalization (\ref{norm}), ${\rm Im}\,K(\omega)$ is used for calculating ${\rm Re}\,K(\omega)$ from (\ref{kkr}), which allows one to calculate Green's function (\ref{Larkin}) for arbitrary {\bf k}. Such obtained spectral functions $A({\bf k}\omega)=-\pi^{-1}{\rm Im}\, G({\bf k}\omega)$ and DOS $\rho(\omega)=N^{-1}\sum_{\bf k}A({\bf k}\omega)$ satisfy the usual normalization conditions with good accuracy.

To check the validity of the used approach its results were compared with data of Monte Carlo simulations carried out in \cite{Groeber} and \cite{Haas}. This comparison is shown in Figs.~\ref{Fig1} and \ref{Fig2}. Here and below $t$ and the intersite distance are set as units of energy and length, respectively. The electron concentration was determined from the relation
$$n=\frac{2}{N}\sum_{\bf k}\int_{-\infty}^\infty\frac{A({\bf k}\omega)}{\exp(\omega/T)+1}\,d\omega,$$
where $T$ was set equal to 0, since in the used approximation equations do not depend on $T$ and actually correspond to $T=0$. Comparing our and Monte Carlo results it should be borne in mind that due to the sign problem the latter were obtained at comparatively high temperature $T=0.5t$. This leads to increased widths of maxima in comparison with zero-temperature results. We did not try to fit widths of maxima in our spectra to Monte Carlo data by an artificial broadening. This was done to demonstrate real spectral shapes, which appear in the present approach. Besides, Monte Carlo spectra for real frequencies were obtained by analytic continuation of imaginary-frequency data using the maximum-entropy method. This method introduces additional inaccuracies in spectral shapes.

Keeping these remarks in mind, from Figs.~\ref{Fig1} and \ref{Fig2} we can conclude that the used approach gives spectral functions in qualitatively and in some cases quantitatively agreement with the Monte Carlo data in the wide range of electron concentrations $0.7\lesssim n\leq 1$ (see also the analogous comparison for the case of half-filling in \cite{Sherman14}). Locations of maxima, general shapes of spectra and their variation with {\bf k} and $n$ are as a rule close. Substantial differences in widths of maxima, observed in some spectra, may be related to the above-mentioned reasons. A similar picture is observed also in the comparison of Monte Carlo \cite{Haas} and our results for $n=0.87$ (not shown here).

Arrows in Figs.~\ref{Fig1} and \ref{Fig2} indicate frequencies of $\delta$-function peaks in the Hubbard-I approximation,
$$\varepsilon_{{\bf k},\pm}=\frac{1}{2}\left(U+t_{\bf k}\pm\sqrt{U^2+t_{\bf k}^2}\right)-\mu.$$
As for the case of half-filling \cite{Sherman14}, one can see that in the used approximation locations of maxima are close to these frequencies.

\begin{figure}[t]
\centerline{\resizebox{0.5\columnwidth}{!}{\includegraphics{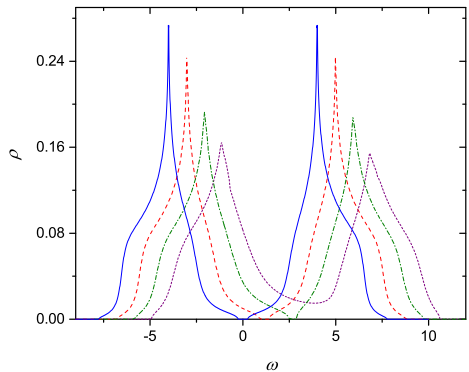}}}
\caption{Densities of states in the $t$-$U$ model for $U=8t$, $n=1$ (blue solid line), $0.99$ (red dashed line), $0.94$ (olive dash-dotted line) and $0.81$ (purple short-dashed line). A 240$\times$240 lattice.} \label{Fig3}
\end{figure}
\begin{figure}[b]
\centerline{\resizebox{0.5\columnwidth}{!}{\includegraphics{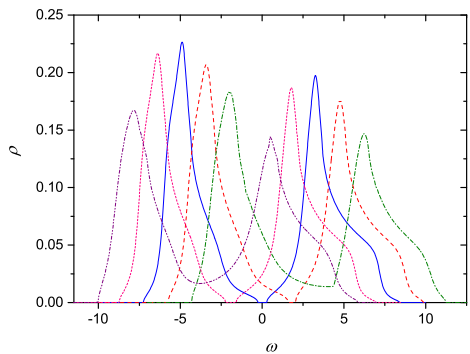}}}
\caption{Densities of states in the $t$-$t'$-$U$ model for $U=8t$, $t'=-0.3t$, $n=1.26$ (purple dash-dot-dotted line), $1.04$ (pink short-dashed line), $1$ (blue solid line), $0.99$ (red dashed line) and $0.87$ (olive dash-dotted line). A 240$\times$240 lattice.} \label{Fig4}
\end{figure}
Widths of spectral maxima are essentially increased away from half-filling. This may be observed in the evolution of the DOS with changing $n$ in Fig.~\ref{Fig3}. Due to the particle-hole symmetry of the $t$-$U$ model only cases $n\leq 1$ are shown. The DOS for $n>1$ can be obtained from the respective case $n<1$ by the specular reflection in the line $\omega=0$. As already noted earlier \cite{Groeber,Sherman15}, with deviation from half-filling the DOS is redistributed in favor of the Hubbard subband, in which the Fermi level is located. The Mott gap decreases monotonously with growing $|1-n|$ and disappears at $|1-n|\approx 0.12$.

\begin{figure}[t]
\centerline{\resizebox{0.5\columnwidth}{!}{\includegraphics{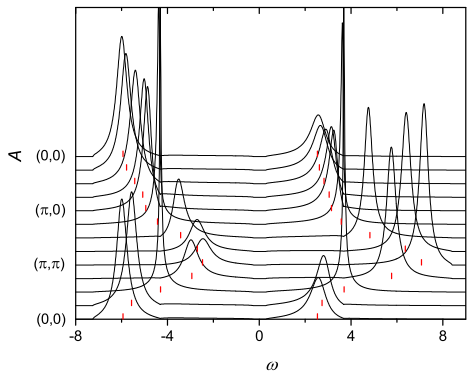}}}
\caption{Spectral functions in the $t$-$t'$-$U$ model for $U=8t$, $t'=-0.3t$, $n=1$ and momenta along the symmetry lines $(0,0)$--$(\pi,\pi)$--$(\pi,0)$--$(0,0)$. A 240$\times$240 lattice. Red pins on the base lines of each curve indicate locations of $\delta$-function peaks in the Hubbard-I approximation.} \label{Fig5}
\end{figure}
\begin{figure}[t]
\centerline{\resizebox{0.5\columnwidth}{!}{\includegraphics{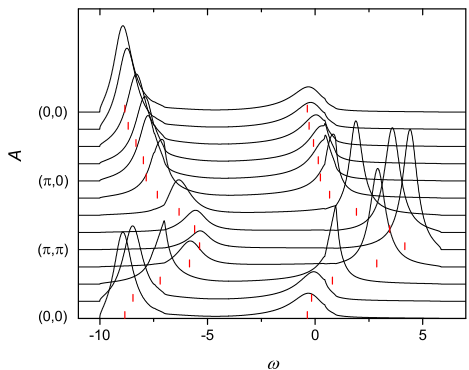}}}
\caption{Same as in Fig.~\protect\ref{Fig5}, but for $n=1.26$.} \label{Fig6}
\end{figure}
Similar behaviour is observed in the DOS of the $t$-$t'$-$U$ model shown in Fig.~\ref{Fig4}. However, in this case the width of the Mott gap reaches its maximum value at $n\approx 1.04$ rather than at $n=1$, as in the $t$-$U$ model. In the $t$-$t'$-$U$ model the maximal gap width is somewhat larger -- $0.64t$ at $U=8t$ and $t'=-0.3t$ in comparison with $0.54t$ at $t'=0$. The dependence of the gap width on $n-1.04$ is strongly asymmetric in the $t$-$t'$-$U$ model. At half-filling, keeping $t'=-0.3t$ and decreasing $U$ we found that the gap disappears at $U_c\approx 7t=7\Delta/8$. This value of the critical repulsion is very close to those found for the semi-elliptical DOS \cite{Hubbard64,Sherman15} and in the $t$-$U$ model \cite{Sherman14}. For the used parameters the positions of maxima in spectral functions of the $t$-$t'$-$U$ model are also close to those in the Hubbard-I approximation, as can be seen in Figs.~\ref{Fig5} and \ref{Fig6}.

\section{Concluding remarks}
In this work, equations for the electron Green's function of the repulsive Hubbard model, derived using the strong coupling diagram technique, were self-consistently solved for different electron concentrations $n$ in the two-dimensional $t$-$U$ and $t$-$t'$-$U$ models. Terms of the first two orders in the expansion in powers of $t/U$ were taken into account in the irreducible part of the equation for Green's function, and the bare internal hopping line in one of these terms was substituted with the dressed one. Comparison of spectral functions, calculated in the $t$-$U$ model for the Hubbard repulsion $U=8t$, with Monte Carlo data shows not only qualitative but in some cases quantitative agreement in positions of maxima. General spectral shapes, their evolution with the wave vector and electron concentration in the wide range $0.7\lesssim n\leq 1$ are also similar. It was found that in the half-filled $t$-$t'$-$U$ model with $t'/t=-0.3$ the Mott transition occurs at $U_c\approx 7\Delta/8$, where $\Delta$ is the initial bandwidth. This critical value is close to those found for the semi-elliptical density of states and in the case $t'=0$. The behaviour of the Mott gap is different in the cases $t'=0$ and $t'=-0.3t$: the maximal gap width is larger in the latter case and is reached at $n=1.04$ rather than at half-filling, as in the case $t'=0$. In both cases positions of spectral maxima are close to those in the Hubbard-I approximation.

The conducted comparison with Monte Carlo results and comparative simplicity of calculations give promise that the strong coupling diagram technique may be a useful tool in consideration of different generalizations of the Hubbard model such as multi-band models used for the description of transition-metal oxides or models containing external fields.

\section*{Acknowledgements}
This work was supported by the research project IUT2-27, the European Regional Development Fund TK114 and the Estonian Scientific Foundation (grant ETF9371).

\end{document}